\documentclass[12pt]{article}
\begin{document}
\large

\begin{center}
\par
 {\bf Schemes of Quark Mixings (Oscillations) and Their Mixing
 Matrices}\\
\vspace{.50cm}
\par
Beshtoev Kh. M. \vspace{0.5cm}
\par
Joint Institute for Nuclear Research, Joliot Curie 6, 141980
Dubna, Moscow region, Russia; Scientific Research Institute
of Applied Mathematics and Automation, KBSC of RAS, Nalchik, Russia  \\
\end{center}

\par
{\bf Abstract}
\par
Three schemes of quark mixings (oscillations) together with their
mixing matrices (analogous to Kabibbo-Kobayashi-Maskawa matrices)
are considered.  In these schemes quark transitions are virtual
since quark masses are different. Two of them belong to the so
called mass mixing schemes (mixing parameters are expressed by
elements of mass matrices) and the third scheme belongs to the
charge mixings one (mixing parameters are expressed through
charges). For these schemes the expressions for transition
probabilities between $d, s, b$ quarks are obtained. The analysis
of situation with the quark mixing parameters in these schemes
is fulfilled.  \\

\vspace*{12pt}
\section{INTRODUCTION}

At present, existence of three following families of leptons and
quarks

$$
\begin{array}{cc} u& \nu _{e}\\ d& e \end{array};\qquad
\begin{array}{cc}c& \nu _{\mu }\\ s& \mu \end{array};\qquad
\begin{array}{cc} t& \nu _{\tau}\\b& \tau \end{array}
\eqno(1)
$$
is established [1]. In the framework of the standard model of weak
interactions [2], i.\,e. at $W$ Ш $Z^{\rm 0}$ boson exchanges,
transitions between different families of leptons or quarks with
flavor number violations do not take place. In the quark sector,
mixings between $d, s, b$ quarks (i.\,e. transitions between
different families of quarks) are described by
Cabibbo--Kobayashi--Maskawa matrices [3]. In works [4], the
dynamical model of transitions between different quark families
(model of dynamical analogy of Cabibbo--Kobayashi--Maskawa
matrices) was proposed. In this model, these transitions are
realized by exchanges of four massive ($B^\pm, C^\pm, D^\pm,
E^\pm$) bosons.

We have a problem with interpretation of the angle mixings in the
standard approach . Consider $K^\pm$, which is produced in strong
interactions, and we want to consider its decay. Since $K$ meson
includes $s$ quark, then when we take into account the weak
interaction, we must use the Cabibbo matrix [3] mixing $s, d$
quarks:
$$
\begin{array}{c}
d_1 =  d\, {\rm cos}\, \theta + s \,{\rm sin}\, \theta,  \\
s_1  =  -d\, {\rm sin}\, \theta  + s\, {\rm cos}\, \theta;
\end{array} \eqno(2)
$$

\noindent i.\,e. $s$ quark transforms in superpositions of $s, d$
quarks:
$$
s  \rightarrow s_1 = -d \,{\rm sin}\, \theta  + s\, {\rm cos}\,
\theta
$$
\par
The matrix element of $K$ meson decay [3] is proportional to ${\rm
sin} \,\theta$, i.\,e. we take into account only the ${\rm
sin}\,\theta$ part from  the above expression, and then the term
proportional to  ${\rm cos}\,\theta$ remains. It means that  only
the part  proportional to ${\rm sin}\,\theta$ decays. However,
from the current experiments we know that $K$ mesons decay fully.
It can happen only if $K$ mesons decay through massive bosons $B$
but not $W$ bosons as it takes place in the model of dynamical
analogy of Cabibbo--Kobayashi--Maskawa matrices [4]. In the
framework of this model, the masses and transition widths of these
bosons were computed and other consequences of quark mixings were
also studied. In the lepton sector, the analogous transitions are
realized by introductions of the same matrices [5].

It is obvious that this problem must be solved in case quark
mixings (or oscillations) take place. Now consider the schemes of
quark mixings (or oscillations) and in the subsequent works we
will return to solution of this problem in the framework of the
suggested approach.

In works [6], three schemes of neutrino mixings (oscillations)
were proposed. The essential difference between quark and lepton
sectors is that quarks are in combined states in hadrons while the
leptons (neutrinos) are in free states. The fact that there are
transitions between neutrinos in free states is an indication on
that in the quark sector the same transitions between quarks will
take place. Besides it is necessary to remark that existence of
hadronic oscillations on examples of $K^{\rm 0}, \bar K^{\rm 0}$
and $B^{\rm 0}, \bar B^{\rm 0}$ oscillations is proven, and these
oscillations are real ones since masses of $K^{\rm 0}$ and $\bar
K^{\rm 0}$ ($B^{\rm 0}$ and $\bar B^{\rm 0}$) are equal. Consider
quark mixings and oscillations in detail.

\section{SCHEMES (TYPES) OF QUARK MIXINGS (OSCILLATIONS) AND THEIR MIXING MATRICES}

In common case there can be two schemes (types) of quark mixings
(oscillations): mass mixing schemes and charge  mixing schemes (as
it takes place in the vector dominance model or vector boson
mixings in the standard model of electroweak interactions).

\subsection{Two Schemes of Quark Mixings (Oscillations) and Their Mixing
Matrices}

In the standard approach [7], it is supposed that quarks (hadrons)
are once created in superposition states, i.\,e. mass matrix is a
nondiagonal one. If  mass matrix is nondiagonal initially, then we
must diagonalize this matrix in order to find eigenstates of
quarks. Then eigenstates are $d_1, s_1, b_1$ quarks (quark mixed
states), i.\,e. $d_1, s_1, b_1$ quarks but not $d , s, b$ quarks
must be created there. It is obvious that it cannot be coordinated
with experimental data. In the strong and weak interactions with
$W$ and $Z^{\rm 0}$ bosons, only $d , s, b$ quarks are created,
i.\,e. initially mass matrix is a diagonal one, and then at
violation of the aromatic numbers this matrix is transformed into
nondiagonal one [6, 8]. We stress this point for its fundamental
importance.
\par
If we work in the framework of the original approach [7], then
these quark transitions (oscillations) must be real transitions
(oscillations), i.\,e. real transitions between quarks must take
place there. It is clear that this supposition violates the law of
energy-momentum conservation [6, 8]. But at $K^{\rm 0}
\leftrightarrow \bar K^{\rm 0}$ transitions, oscillations are real
since masses of $d$ and $\bar d$, $s$ and $\bar s$ are equal. But
at transitions between different-mass hadrons ($\pi^\pm
\leftrightarrow K^\pm$), these transitions will be virtual [4, 9].

We can also see that there are two cases of quark transitions
(oscillations) in the scheme of mass mixings by analogy with the
neutrino transitions (oscillations)~[9]. \\

\subsubsection{Use of the Corrected Standard
Scheme of Neutrino Mixings (Oscillations) for Consideration of the
Quark Mixings (Oscillations)}

\par
 We can use the corrected standard scheme [6]
of neutrino mixings (oscillations) for consideration of the quark
mixings (oscillations) since quarks as well as neutrinos are also
fermions. The following corrected standard scheme belongs to the
so-called mass mixing scheme since mixing parameters are expressed
through elements of mass matrix. Since the quarks are created in
the strong interactions where the aroma numbers are conserved then
$d, s$ quark mass matrix originally must have diagonal form (for
simplification we consider two quark mixings)
$$
\left(\begin{array}{cc} m_{d} & 0\\
0 & m_{s}\end{array}\right),
$$
then for presence of the weak
interactions violating aroma numbers $d, s$ quark mass get the
following non diagonal form:

\setcounter{equation}{2}
\begin{equation}
\left(\begin{array}{ccc}m_{d} & m_{d s}\\
m_{s d} & m_{s}\end{array}\right),
\end{equation}
Diagonalizing this matrix
\begin{equation}
\left(\begin{array}{cc} m_{d_1} & 0\\
0 & m_{s_1}\end{array}\right),
\end{equation}
we pass to intermediate quark states- $d_1, s_1$ and then
$$
\begin{array}{c}
d = {\rm cos}\, \theta  d_1 - {\rm sin}\, \theta s_1,    \nonumber  \\
s = {\rm sin}\, \theta  d_1 + {\rm cos}\, \theta  s_1.
\end{array}
$$
 In this case, the probability of $d \rightarrow s$ transition
(oscillation) is described by the following expression:
\begin{equation}
P(d \rightarrow s, t) =  {\rm sin}^2\, 2\theta\, {\rm sin}^2
\left[\pi t\frac{\mid m_{d_1}^2 - m_{s_1}^2 \mid}{2 p_{d}}
\right],
\end{equation}
where $p_{d}$ is a momentum of $d$ quark,
\begin{equation}
sin^2\, 2\theta = \frac{4m^2_{d, s}}{(m_{d} - m_{s})^2 + 4m^2_{d,
s}},
\end{equation}
\noindent and
\begin{equation}
m_{d_1, s_1} = {1\over 2} \left[ (m_{d} + m_{s}) \pm \left((m_{d}
- m_{s})^2 + 4 m^{2}_{d s} \right)^{1/2} \right].
\end{equation}
At these transitions (oscillations), quarks remain on their mass
shell and these transitions (oscillations) must be virtual.

It is interesting to remark that expression (5) can be obtained
from the Breit-Wigner distribution [11]
\begin{equation}
P \sim \frac{(\Gamma/2)^2}{(E - E_0)^2 + (\Gamma/2)^2}
\end{equation}
by using the following substitutions:
\begin{equation}
E = m_{d},\hspace{0.2cm} E_0 = m_{s},\hspace{0.2cm} \Gamma/2 =
2m_{d, s},
\end{equation}
where $\Gamma/2 \equiv W(... )$ is a width of $d \rightarrow s$
transition, then we can use a standard method [10, 12] for
calculating this value. Then the probability of $d \rightarrow s$
transitions is defined by these quark masses and widths of their
transitions.
\par
Expression for length of these oscillations has the following
form:
\begin{equation}
L_{o} = 2\pi  {2p_d \over {\mid m^{2}_{s_1} - m^{2}_{d_1} \mid}}.
\end{equation}

Above, we considered the case of two quark transitions
(oscillations). In common case, we must consider three quark
transitions (oscillations). For a complete description of three
quark oscillations we must have six parameters (we suppose that
this mass matrix is symmetric in respect to the diagonal one),
\begin{equation}
\left(\begin{array}{ccc}m_{d} & m_{d s} & m_{d b} \\
m_{s d} & m_{s} & m_{s b}\\
m_{d b} &  m_{s b} & m_{b} \end{array}\right),
\end{equation}
three diagonal terms of this matrix are masses of three physical
quarks $m_{d}, m_{s}, m_{b}$, and three nondiagonal mass terms of
this matrix are $m_{d s}$, $m_{d b}, m_{s b}$-quarks transition
widths. Since in the expression for quark transition probabilities
the squared mass differences are used in reality, we need only
five parameters (for further simplification, physical quark masses
are used). Besides, if mass matrix is complex, there appears one
parameter connected with $CP$ violation.

These mixing angles can be connected with the
Cabibbo--Kobayashi--Maskawa mixing matrix $V$ [3]. We will choose
a parameterization of  the mixing matrix $V$ in the form proposed
by Maiani [13]: \setcounter{equation}{11}
\begin{equation}
{V = \left( \begin{array} {ccc}1& 0 & 0 \\
0 & c_{\gamma} & s_{\gamma} \\ 0 & -s_{\gamma} & c_{\gamma} \\
\end{array} \right) \left( \begin{array}{ccc} c_{\beta} & 0 &
s_{\beta} \exp(-i\delta) \\ 0 & 1 & 0 \\ -s_{\beta} \exp(i\delta)
& 0 & c_{\beta} \end{array} \right) \left( \begin{array}{ccc}
c_{\theta} & s_{\theta} & 0 \\ -s_{\theta} & c_{\theta} & 0 \\ 0 &
0 & 1 \end{array}\right)} ,
\end{equation}

$$
c_{d s} = \cos {\theta }, \quad s_{d s} =\sin{\theta}, \quad
c^2_{d s} + s^2_{d s} = 1;
$$
$$
c_{d b} = \cos {\beta }, \quad s_{d b} =\sin{\beta}, \quad c^2_{d
b} + s^2_{d b} = 1; \eqno(12')
$$
$$
c_{s b} = \cos {\gamma} , \quad s_{s b} =\sin{\gamma}, \quad
c^2_{s b} + s^2_{s b} = 1;
$$
$$
 \exp(i\delta) = \cos{\delta } + i \sin{\delta}.
$$
In our approximation, the value of $\delta$ can be considered
equal to zero.

Equations for mixing angles expressed through elements of mass
matrix has the following form:
$$
s_{d s} = \sin {\theta } = \frac{1}{\sqrt{2}} \left[ 1 -
\frac{\mid m_{s} - m_{d} \mid}{\sqrt{(m_{s} - m_{d})^2 + (2 m_{d
s})^2}} \right], \eqno(13)
$$

$c^2_{d s} = 1 - s^2_{d s}$;
$$
s_{d b} = \sin {\beta } = \frac{1}{\sqrt{2}} \left[ 1 - \frac{\mid
m_{b} - m_{d} \mid}{\sqrt{(m_{b} - m_{d})^2 + (2 m_{d b})^2}}
\right], \eqno(14)
$$

$c^2_{d b} = 1 - s^2_{d b}$;
$$
s_{s b} = \sin {\gamma} = \frac{1}{\sqrt{2}} \left[ 1 - \frac{\mid
m_{b} - m_{s} \mid}{\sqrt{(m_{b} - m_{s})^2 + (2 m_{s b})^2}}
\right], \eqno(15)
$$

$c^2_{s b} = 1 - s^2_{s b}$. \\

{\bf Analysis of Status of Quark Mixing Parameters in the Scheme
of Mass Mixings} \\

\par
With this aim we use the following data on mixing angles obtained
in the framework of Cabibbo--Kobayashi--Maskawa matrices [1]:
\setcounter{equation}{15}
$$
\begin{array}{c}
1)\quad tg \,\theta \cong {\rm sin}\, \theta = 0.218\div0.224;\\
2)\quad tg\, \beta  \cong {\rm sin} \,\beta = 0.032\div0.054;\\
3)\quad tg\, \gamma  \cong  {\rm sin}\, \gamma = 0.002\div0.007.
\end{array}
\eqno(16)
$$
Expressions for squared mass differences and their expansions have
the following form $(d_1 \to 1, s_1 \to 2, b_1 \to 3)$:
$$
\Delta m^2_{2 1} = m^2_2 - m^2_1 = (m_{s} + m_{d}) \sqrt{(m_{s} -
m_{d})^2 + (2 m_{d s})^2} \eqno(17)
$$
if $ 2 m_{d s} \gg \mid m_{s} - m_{d} \mid $, then
$$
\Delta m^2_{2 1} = (m_{s} + m_{d}) 2 m_{d s} \left[1 +
\frac{(m_{s} - m_{d})^2}{2(2 m_{d s})^2}\right],\eqno(17')
$$
and if $ 2 m_{d s} \ll \mid m_{s} - m_{d} \mid $, then
$$
\Delta m^2_{2 1} = (m^2_{s} - m^2_{d}) \left[1 + \frac{(2 m_{d
s})^2} {2(m_{s} - m_{d})^2}\right]; \eqno(17'')
$$
$$
\Delta m^2_{3 1} = m^2_3 - m^2_1 = (m_{b} + m_{d}) \sqrt{(m_{b} -
m_{d})^2 + (2 m_{d b})^2}, \eqno(18)
$$
if  $2 m_{d b} \gg \mid m_{b} - m_{d} \mid $, then
$$
\Delta m^2_{3 1} = (m_{b} + m_{d})2 m_{d b} \left[1 + \frac{(m_{b}
- m_{d})^2}{2 (2 m_{d b})^2} \right], \eqno(18')
$$
and if  $2 m_{d b} \ll \mid m_{b} - m_{d} \mid $, then
$$
\Delta m^2_{3 1} = (m^2_{b} - m^2_{d}) \left[1 + \frac{ (2 m_{d
b})^2}{2(m_{b} - m_{d})^2}\right]; \eqno(18'')
$$
$$
\Delta m^2_{3 2} = m^2_3 - m^2_2 = (m_{b} + m_{s}) \sqrt{(m_{b} -
m_{s})^2 + (2 m_{s b})^2} \eqno(19)
$$
if $2 m_{s b} \gg \mid m_{b} - m_{s} \mid $, then
$$
\Delta m^2_{3 2} = (m_{b} + m_{s})2 m_{s b} \left[1 + \frac{(m_{b}
- m_{s})^2}{2 (2 m_{s b})^2} \right], \eqno(19')
$$
and if $2 m_{s b} \ll \mid m_{b} - m_{s} \mid $, then
$$
\Delta m^2_{3 2} = (m^2_{b} - m^2_{s}) \left[1 + \frac{(2 m_{s
b})^2}{2(m_{b} - m_{\nu_mu})^2} \right]. \eqno(19'')
$$
\par
The current masses of $d, s, b$ quarks are [1]:
$$m_d \simeq 3\div 9\, {\rm MeV},$$
$$m_s \simeq 60 \div 170 \,{\rm MeV},
\eqno(20)
$$
$$m_b \simeq 4.0 \div 4.5\, {\rm GeV}.$$
\par
Now using values of these masses we turn to consideration of the
situation with quark mixings (oscillations).
\par
For $d, s$ quarks we have (by diagonalization of mass matrix)
$$
\begin{array}{c} {\rm sin}\, 2\theta = \frac{2 m_{ds}}{\sqrt{(m_d - m_s)^2 + 4
m_{d s}^2}}, \\
 P(d \rightarrow s, t) =  {\rm sin}^2\, 2\theta\, {\rm sin}^2 \left[\pi
t\frac{\mid m_{d_1}^2 - m_{s_1}^2 \mid}{2 p_{d}} \right],
\end{array} \eqno(21)
$$
$$
\begin{array}{c} d_1 =  d \,{\rm cos}\, \theta + s\, {\rm sin}\, \theta,  \\
s_1  =  -d \sin \theta  + s\, {\rm cos} \theta. \end{array}
\eqno(22)
$$
\par
For $d, b$ quarks we have (by diagonalization of mass matrix)
$$
\begin{array}{c}
 {\rm sin}\, 2\beta = \frac{2 m_{d b}}{\sqrt{(m_d - m_b)^2 + 4
m_{d b}^2}}, \\[4mm]
P(d \rightarrow b, t) =  {\rm sin}^2\, 2\theta\, {\rm sin}^2
\left[\pi t\frac{\mid m_{d_1}^2 - m_{b_1}^2 \mid}{2 p_{d}}
\right],
\end{array} \eqno(23)
$$
$$
\begin{array}{c} d_1 =  d \,{\rm cos}\, \beta + b\, {\rm sin}\, \beta,  \\
b_1  =  -d \,\sin \beta  + b\, {\rm cos}\, \beta. \end{array}
\eqno(24)
$$
\par
For $s, b$ quarks we have (by diagonalization of mass matrix)
$$
\begin{array}{c} {\rm sin}\, 2\gamma = \frac{2 m_{s b}}{\sqrt{(m_s - m_b)^2 + 4
m_{s b}^2}}, \\[4mm]
P(s \rightarrow b, t) =  {\rm sin}^2\, 2\theta\, {\rm sin}^2
\left[\pi t\frac{\mid m_{s_1}^2 - m_{b_1}^2 \mid}{2 p_{s}}
\right],
\end{array} \eqno(25)
$$
$$
\begin{array}{c} s_1 =  s \,{\rm cos}\, \gamma + b\, {\rm sin}\, \gamma,  \\
b_1  =  -s \sin \,\gamma  + b \,{\rm cos}\, \gamma. \end{array}
\eqno(26)
$$
If we use angle values from (16) and quark masses from (20), then
we can rewrite expressions (21), (23) and (25) in the following
form:
$$
{\rm sin}\, 2\theta \simeq \frac{2 m_{d s}}{ m_s}, \quad m_{d s}
\simeq \frac{1}{2} m_s \,{\rm sin}\, 2 \theta\, \simeq
16.0\div16.35 \,{\rm MeV}, \eqno(27)
$$
$$
{\rm sin}\, 2\beta \simeq \frac{2 m_{db}}{ m_b} , \quad m_{d b}
\simeq \frac{1}{2} m_b\,{\rm  sin}\, 2 \beta \,\simeq
172.0\div193\,{\rm  MeV}, \eqno(28)
$$
$$
{\rm sin}\, 2\gamma \simeq \frac{2 m_{sb}}{ m_b} , \quad m_{s b}
\simeq \frac{1}{2} m_b\, {\rm sin}\, 2\gamma \,\simeq 18\div20.2
\,{\rm MeV}. \eqno(29)
$$
In this approach we interpret nondiagonal mass terms of mass
matrix as transition widths between quarks. Then it is not clear
how this value can be more than quark mass value as it takes place
in (27). In any case such enormous values of widths can arise only
if $s, b$ quarks are resonance states. Obviously these resonance
can originate only outside the  standard weak interactions (see
below).

If we compute  values of nondiagonal mass terms (quark transition
widths) of mass matrix in the framework of the standard weak
interactions, then using Eq. (21) we get
$$
m_{ds} \simeq {\rm sin}\, \theta m_s. \eqno(30)
$$

It is interesting to compute this angle mixing in the standard
model in the framework of some consistent supposition on the
analogy of $K^{\rm 0}, \bar K^{\rm 0}$ or $\pi\pm, K\pm$ mixings
[9]. To do it, we suppose that $d \leftrightarrow s$ transitions
are generated through exchange of massive boson $W'$. Then,
formally, we can get

$$
m_{ds} \simeq 2 W(d \rightarrow s) \simeq
$$
$$
\simeq (G_F)^2 {{f'}^2_\pi m_s^3\over 8\pi} ({m_W\over m_{W'}})^4
= m_s\, {\rm sin}\,\theta'. \eqno(31)
$$
\par
Even if we take $m_{W'} \simeq m_W$ and ${f'}_\pi \sim$ a few GeV,
we come to the following result:
$$
{\rm sin}\, \theta' = (G_F)^2 {{f'}^2_\pi m_s^2\over 8\pi} \ll
{\rm sin}\,\theta \simeq \sqrt{0.048}. \eqno(32)
$$
So, we see that the angle mixing ${\rm sin}\,\theta'$ obtained in
the standard method is a very small value and much less than ${\rm
sin}\,\theta$ in the Cabibbo--Kobayashi--Maskawa matrices. It is
clear that we cannot obtain fit values for mixing angles in this
approach. Probably we must suppose that there must be a new
left--right symmetrical interaction, which can generate masses of
quarks, and moreover some quarks must be resonances of this
interaction (by analogy with the strong interactions).

Impossibility of obtaining values for quark transition widths,
which are of the order of $m_{d s}, m_{d b}, m_{s b}$ in Eqs.
(27)--(29), in the framework of the weak interactions (see (32))
is an indication that the mass mixing schemes cannot fit the
description of quark mixings. Unfortunately the same situation can
take place in the neutrino mixing cases although this approach is
used everywhere in description of experiments on neutrino mixings
and oscillations. \\

\subsubsection{The Case of Quark Mixings without Mass Shell
Changing}

\par
Above we considered the case when virtual quark transitions take
place with change of quark masses. Another case is also possible,
when $d$ quark transits into $s$ quark without changing mass,
i.\,e. $m^{*}_{s} = m_{d}$, then
$$
{\rm tg}\, 2\theta = \infty, \eqno(33)
$$
$\theta = \pi/4$, and
$$
{\rm sin}^2\, 2\theta = 1. \eqno(34)
$$
\par
In this case, the probability of the $d \rightarrow s$ transition
(oscillation) is described by the following expression:
$$
P(d \rightarrow s, t) = {\rm sin}^2 \left[\pi t\frac{4 m_{d,
s}^2}{2 p_d} \right ]. \eqno(35)
$$

Expression for length of oscillations in this case has the
following form:
$$
L_{o} = 2\pi  \frac{2p_d} {(2m_{d s})^2}.
$$

In order to make these virtual oscillations real, their
participation in quasi-elastic interactions is necessary for their
transitions to their own mass shells [10].

The Kobayashi--Maskawa-type matrix in this case is a trivial one,
and it has the following form:
$$
{V = \left( \begin{array} {ccc}1& 0 & 0 \\
0 & c_{\gamma} & s_{\gamma} \\ 0 & -s_{\gamma} & c_{\gamma} \\
\end{array} \right) \left( \begin{array}{ccc} c_{\beta} & 0 &
s_{\beta} \exp(-i\delta) \\ 0 & 1 & 0 \\ -s_{\beta} \exp(i\delta)
& 0 & c_{\beta} \end{array} \right) \left( \begin{array}{ccc}
c_{\theta} & s_{\theta} & 0 \\ -s_{\theta} & c_{\theta} & 0 \\ 0 &
0 & 1 \end{array}\right)}, \eqno(36)
$$
\par
$$
c_{e \mu} = \cos {\theta } = \frac{1}{\sqrt2}, \quad s_{e \mu}
=\sin{\theta}= \frac{1}{\sqrt2};
$$
$$
c_{e \tau} = \cos {\beta }= \frac{1}{\sqrt2}, \quad s_{e \tau}
=\sin{\beta}= \frac{1}{\sqrt2}; \eqno(37)
$$
$$
c_{\mu \tau} = \cos {\gamma}= \frac{1}{\sqrt2}, \quad s_{\mu \tau}
=\sin{\gamma}= \frac{1}{\sqrt2};
$$
$$
 \exp(i\delta) = 1.
$$
In our approximation, the value of $\delta$ can be considered to
be equal to zero.

In this case
$$
{\rm sin}^2\, 2\theta = {\rm sin}^2\, 2\beta = {\rm sin}^2\,
2\gamma = 1, \eqno(38)
$$
we have
$$
\Delta m^2_{21} = (2m_{d s})^2,
$$
$$
\Delta m^2_{31} = (2m_{d b})^2, \eqno(39)
$$
$$
\Delta m^2_{32} = (2m_{d s})^2.
$$

It is necessary to remark that in physics all the processes are
realized through dynamics. Unfortunately, in this mass mixing
scheme the dynamics is absent. Probably that is  an indication of
the fact that these schemes are incomplete ones, i.\,e. these
schemes demand a physical substantiation (see Sec. 2.2).

In principle we cannot exclude this type of quark mixings since
lengths of quark transitions (oscillations) in this case are much
more than it were in previous case; therefore on the background of
previous transitions it is hard to observe these transitions.

Obviously, these schemes will work only if quark oscillations take
place in reality (it is clear that there also can be
quark mixings in absence of quark oscillations).\\

\vspace{-12pt}
\subsection{The Scheme of Quark Mixings (Oscillations) via Charges}

The third scheme (type) of mixings or transitions between quarks
can be realized by mixings of the quark fields by analogy with the
vector dominance model ($\gamma-\rho^{\rm 0}$) and $Z^{\rm
0}-\gamma$ mixings as it takes place in the particle physics [2,
14]. Then in the case of two quarks, we have
\setcounter{equation}{39}
\begin{equation}
\begin{array}{c}
q_1 = {\rm cos}\, \theta d + {\rm sin}\, \theta s, \\
q_2 = -{\rm sin}\, \theta d + {\rm cos}\, \theta s. \
\end{array}
\end{equation}
In the case of three quarks, we can also choose parameterization
of the mixing matrix $V$ in the form proposed by Maiani [13]:
$$
{V = \left( \begin{array} {ccc}1& 0 & 0 \\
0 & c_{\gamma} & s_{\gamma} \\ 0 & -s_{\gamma} & c_{\gamma} \\
\end{array} \right) \left( \begin{array}{ccc} c_{\beta} & 0 &
s_{\beta} \\ 0 & 1 & 0 \\ -s_{\beta} & 0 & c_{\beta}
\end{array} \right) \left( \begin{array}{ccc} c_{\theta} &
s_{\theta} & 0 \\ -s_{\theta} & c_{\theta} & 0 \\ 0 & 0 & 1
\end{array}\right)}; \eqno(41)
$$
$$
c_{e \mu} = \cos {\theta } \quad s_{e \mu} =\sin{\theta},
 \quad c^2_{e \mu} + s^2_{e \mu} = 1;
$$
$$
c_{e \tau} = \cos {\beta }, \quad s_{e \tau} =\sin{\beta}, \quad
c^2_{e \tau} + s^2_{e \tau} = 1; \eqno(42)
$$
$$
c_{\mu \tau} = \cos {\gamma}, \quad s_{\mu \tau} =\sin{\gamma},
\quad c^2_{\mu \tau} + s^2_{\mu \tau} = 1.
$$

The charged current in the standard model of weak interactions for
two quark families has the following form:
$$
j^\alpha  = \left(\begin{array}{cc} \bar u \bar c
\end{array}\right)_L \gamma^\alpha V \left(\begin{array}{c} d \\
s \end{array} \right)_L,
$$
$$
V = \left(\begin{array}{cc} \cos \theta & \sin \theta \\
-\sin \theta & {\rm cos}\, \theta \end{array}\right), \eqno(43)
$$
and then the interaction Lagrangian is
$$
{\cal L} = \frac{g}{\sqrt{2}} j^\alpha W^{+}_\alpha  + h.c.
\eqno(44)
$$
and
$$
\begin{array}{c}
d = {\rm cos}\, \theta  q_1 - {\rm sin}\, \theta q_2,           \\
s = {\rm sin}\, \theta q_1 + {\rm cos}\, \theta q_2.
\end{array}
\eqno(45)
$$
Then, taking into account that the charges of $q_1, q_2$ quarks
are $g_1, g_2$, we get
$$
g\, {\rm cos}\, \theta = g_1, \quad g\, {\rm sin}\, \theta = g_2,
\eqno(46)
$$
i.\,e.
$$
{\rm cos}\, \theta = \frac{g_1}{g}, \quad {\rm sin}\, \theta =
\frac{g_2}{g}. \eqno(47)
$$
Since ${\rm sin}^2 \,\theta + {\rm cos}^2 \,\theta = 1$, then
$$
g = \sqrt{g_1^2 + g_2^2}
$$
and
$$
{\rm cos}\, \theta = \frac{g_1}{\sqrt{g_1^2 + g_2^2}}, \quad {\rm
sin}\, \theta = \frac{g_2}{\sqrt{g_1^2 + g_2^2}}. \eqno(48)
$$
\par
If we suppose that $ g_1 \cong g_2 \cong \frac{g}{\sqrt{2}}$, then
$$
{\rm cos}\, \theta \cong  {\rm sin}\, \theta \cong
\frac{1}{\sqrt{2}}. \eqno(49)
$$
It is not difficult to turn to consideration of the case of three
quark types $d, s, b$. Since the weak couple constants $g_d, g_s,
g_b$ of $d, s, b$ quarks are approximately equal in reality,
i.\,e. $g_{d}\simeq g_s \simeq g_b$, then the angle mixings are
nearly maximal: \setcounter{equation}{49}
$$
\begin{array}{c}
 {\rm cos}\, \theta = {\rm cos}\, \theta_{d s} \cong  {\rm sin}\, \theta_{d s} \cong
\frac{1}{\sqrt{2}}, \\ {\rm cos}\, \beta = {\rm cos}\, \theta_{d
b} \cong  {\rm sin}\, \theta_{d b} \cong
\frac{1}{\sqrt{2}}, \\
{\rm cos}\, \gamma = {\rm cos}\, \theta_{s b} \cong  {\rm sin}\,
\theta_{s b} \cong \frac{1}{\sqrt{2}}.
\end{array}
\eqno(50)
$$
In expression (16), experimental data for quark mixing angles were
given. These values are in serious discrepancy with the same
values in (50). These discrepancy can be eliminated if we suppose
that quark charges (or couple constants) of the interactions,
which violate quark aromatic numbers, are different from the weak
charges (couple constants) of $d, s, b$ quarks. Then we can use
Eq. (46)--(47) for determination of $q_1, q_2, q_3$ quark charges
by using values from Eq. (16): \setcounter{equation}{50}
$$
\begin{array}{c}
g'_1 = g\, {\rm cos}\,  \theta,\,\,\,\,\, g'_2 = g\, {\rm sin}\,  \theta,\\
g''_1 = g \,{\rm cos}\,  \beta, \,\,\,\, g''_3 = g\, {\rm sin}\, \beta, \\
g'''_2 = g\, {\rm cos }\,\gamma,\,\,\,\,\, g'''_3 = g\, {\rm
sin}\, \gamma. \end{array} \eqno(51)
$$
It is also possible to use expression (48) as an independent one
and use it for determination of $q_1, q_2, q_3$ couple constants
(it is consequence of normalization conservation and then there
can be no connections between these couple constants and the
above-given quark couple constants).

As it is stressed above, in the case of mass mixing scheme we have
no dynamical basing in contrast to the case of charge mixing
scheme, but these schemes may be jointed if quark masses have the
following form:
$$
m_{q_i} = g_i v, \quad i = d, s, b, \eqno(52)
$$
where $v$ is constant type of constant in the Higgs mechanism
[15]. And then the problem of dynamical substantiation in this
scheme is solved. The problem of using the mass mixing schemes for
description of quark mixings (oscillations) is also solved, but
now nondiagonal terms of quark mass matrix cannot be interpreted
as quark transition widths.

\section{CONCLUSION}

Unfortunately, we do not know concrete mechanism of quark mixings
or oscillations; therefore, it is necessary to consider all
realistic schemes of quark mixings and oscillations. In this work,
three schemes of quark mixings (oscillations) together with their
mixing matrices (analogous to Cabibbo--Kobayashi--Maskawa
matrices) were considered. In these schemes, quark transitions are
virtual since quark masses are different. Two of them belong to
the so-called mass mixing schemes (mixing parameters are expressed
by elements of mass matrices), and the third scheme belongs to the
charge mixing ones (mixing parameters are expressed through
charges). For these schemes the expressions for transition
probabilities between $d, s, b$ quarks were obtained. The analysis
of situation with the quark mixing parameters in these schemes was
fulfilled. It was shown that in principle it is impossible to
obtain values for quark transition widths given in Eqs. (27)--(29)
in the framework of the weak interactions (see (32)). It is an
indication that the mass mixing schemes cannot fit for the
description of quark mixings (oscillations), although if quark
mass origin has a Higgs nature (see Eq. (52)) then this problem is
solved. In this case, we cannot consider the nondiagonal mass
terms of quark mass matrix as quark transition widths any longer.

So, expressions (4)--(9), (13)--(15), (21)--(26), (35)--(38),
(50), (51) can be used for interpretation of experimental data on
quark mixings and oscillations. These quark mixings and
oscillations will be manifested as mixings and oscillations of
hadrons.


\begin{thebibliography}{100}
\small{

\bibitem{1}
Review of Part. Prop. // Phys. Rev. D. 2002. V.\,66, No.\,1.
P.\,010001.

\bibitem{2}
{\it Glashow S.\,L.}~// Nucl. Phys. 1961. V.\,22. P.\,579;\\
{\it Weinberg S.}~// Phys. Rev. Lett. 1967. V.\,19. P.\,1264;\\
{\it Salam A.}~// Proc. of the 8th Nobel Symp. / Ed. N.
Svarthholm. Almgvist and Wiksell, Stockholm, 1968. P.\,367.

\bibitem {3}
{\it  Cabibbo N.}~// Phys. Rev. Lett. 1963. V.\,10. P.\,531;\\
{\it Kobayashi M., Maskawa K.}~// Prog. Theor. Phys. 1973. V.\,49.
P.\,652.

\bibitem {4}
{\it Beshtoev Kh.\,M.} JINR, E2-94-293. Dubna,1994;
Turkish Journ. of Phys. 1996. V.\,20. P.\,1245;\\
{\it Beshtoev Kh.\,M.} JINR Commun. E2-95-535. Dubna, 1995;\\
{\it Beshtoev Kh.\,M.} JINR Commun. А2-96-450. Dubna, 1996;\\
{\it  Beshtoev Kh.\,M.} JINR Commun. E2-97-210. Dubna, 1997;\\
{\it Beshtoev Kh.\,M.} Hadronic J. 2000. V.\,23. P.\,295.

\bibitem{5}
{\it Beshtoev Kh.\,M.} JINR Commun. E2-91-183. Dubna, 1991; Proc.
of III  Int. Symp. on Weak and Electromag. Int. in Nucl.
Singapoure, 1992. P.\,781; 13th European Cosmic Ray Symp. CERN,
Geneva. HE-5-13.


\bibitem{6}
{\it  Beshtoev Kh.\,M.} JINR Commun. E2-92-318. Dubna, 1992;\\
{\it Beshtoev Kh.\,M.}~// Proc. of 27th Intern. Cosmic Ray
Conf. Germany, 2001. V.\,4. P.\,1186;\\
{\it Beshtoev Kh.\,M.}~// Proc. of 28th Intern. Cosmic Ray
Conf. Japan, 2003. V.\,1. P.\,1503.\\
{\it Beshtoev Kh.\,M.} JINR Commun. E2-2004-58. Dubna, 2004.

\bibitem{7}
{\it  Bilenky S.\,M., Pontecorvo B.\,M.}~// Phys. Rep. C. 1978.
V.\,41. P.\,225;\\
{\it Boehm F., Vogel P.}~// Physics of Massive Neutrinos:
Cambridge
Univ. Press, 1987;\\
{\it Bilenky S.\,M., Petcov S.\,T.}~// Rev. of Mod. Phys. 1977.
V.\,59. P.\,631.

\bibitem{8}
{\it  Beshtoev Kh.\,M.} JINR Commun. E2-2003-155. Dubna, 2003.

\bibitem{9}
{\it  Beshtoev Kh.\,M.} JINR Commun. E2-99-137. Dubna, 1999;\\
{\it Beshtoev Kh.\,M.} JINR Commun. E2-2000-229. Dubna, 2000.

\bibitem{10}
{\it  Beshtoev Kh.\,M.} hep-ph/9911513;
 Hadronic J.  2000. V.\,23. P.\,477;
Proc. of 27th Int. Cosmic Ray Conf. Germany, Hamburg, 7--15 August
2001. V.\,3. P.\,1186.

\bibitem{11}
{\it  Blatt J.\,M., Weisskopf V.\,F.} The Theory of Nuclear
Reactions. INR T\,R 42.

\bibitem{12}
{\it  Beshtoev Kh.\,M.} JINR Commun. E2-99-307. Dubna, 1999; JINR
Commun. E2-99-306. Dubna, 1999.

\bibitem{13}
  {\it  Maiani L.}~// Proc. Int. Symp. on Lepton--Photon
Int. Hamburg, DESY. P.\,867.

\bibitem{14}
{\it Beshtoev Kh.\,M.} Preprint INR of Academy of Sciences of
USSR. P-217. Moscow, 1981.

\bibitem{15}
{\it  Higgs P.\,W.}~// Phys. Lett. 1964. V.\,12. P.\,132;
Phys. Rev. 1966. V.\,145. P.\,1156;\\
{\it Englert F., Brout R.}~// Phys. Rev. Lett. 1964. V.\,13. P.\,321;\\
{\it Guralnik G.\,S., Hagen C.\,R.,  Kible T.\,W.\,B. }~// Phys.
Rew. Lett. 1964. V.\,13. P.\,585.

}

\end{thebibliography}
\end{document}